\documentclass[apj,revtex4]{emulateapj_2013}

\usepackage{amsmath,amssymb,pdflscape,mathptmx,graphicx,natbib,rotating,rotfloat,multirow}
\usepackage[breaklinks,colorlinks,citecolor=blue,linkcolor=blue,urlcolor=blue]{hyperref}

\def\hst{\textit{HST}}
\newcommand\ltsim{\raisebox{-.5ex}{$\;\stackrel{<}{\sim}\;$}}

\newcommand\na{{NewA}}%
\newcommand\pasa{{PASA}}%

\newcommand{\CCPP}{CCPP, New York University, 4 Washington Place, New York, NY 10003, USA}
\newcommand{\AMNH}{Department of Astrophysics, American Museum of Natural History, New York, NY 10024, USA}
\newcommand{\JHU}{Department of Physics and Astronomy, The Johns Hopkins University, Baltimore, MD 21218, USA}
\newcommand{\STSCI}{Space Telescope Science Institute, Baltimore, MD 21218, USA}
\newcommand{\ANU}{Research School of Astronomy and Astrophysics, The Australian National University, Cotter Road, Weston Creek, ACT 2611, Australia}
\newcommand{\CAASTRO}{ARC Centre of Excellence for All-Sky Astrophysics (CAASTRO), Australian National University, Canberra, ACT 2611, Australia}

\shorttitle{Discovery of $^{57}$Co in SN 2012cg}
\shortauthors{Graur et al.}

\begin{document}

\title{Late-Time Photometry of Type Ia Supernova SN 2012cg Reveals the Radioactive Decay of $^{57}$Co}

\author{Or~Graur\altaffilmark{1,2},
David~Zurek\altaffilmark{2},
Michael~M.~Shara\altaffilmark{2},
Adam G. Riess\altaffilmark{3,4},
Ivo R. Seitenzahl\altaffilmark{5,6}, and 
Armin Rest\altaffilmark{4}
}

\altaffiltext{1}{\CCPP}
\altaffiltext{2}{\AMNH}
\altaffiltext{3}{\JHU}
\altaffiltext{4}{\STSCI}
\altaffiltext{5}{\ANU}
\altaffiltext{6}{\CAASTRO}

\email{orgraur@nyu.edu}




\begin{abstract}
\noindent Seitenzahl et al. have predicted that roughly three years after its explosion, the light we receive from a Type Ia supernova (SN Ia) will come mostly from reprocessing of electrons and X-rays emitted by the radioactive decay chain $^{57}{\rm Co}~\to~^{57}{\rm Fe}$, instead of positrons from the decay chain $^{56}{\rm Co}~\to~^{56}{\rm Fe}$ that dominates the SN light at earlier times. Using the {\it Hubble Space Telescope}, we followed the light curve of the SN Ia SN 2012cg out to $1055$ days after maximum light. Our measurements are consistent with the light curves predicted by the contribution of energy from the reprocessing of electrons and X-rays emitted by the decay of $^{57}$Co, offering evidence that $^{57}$Co is produced in SN Ia explosions. However, the data are also consistent with a light echo $\sim14$ mag fainter than SN 2012cg at peak. Assuming no light-echo contamination, the mass ratio of $^{57}$Ni and $^{56}$Ni produced by the explosion, a strong constraint on any SN Ia explosion model, is $0.043^{+0.012}_{-0.011}$, roughly twice Solar. In the context of current explosion models, this value favors a progenitor white dwarf with a mass near the Chandrasekhar limit.
\end{abstract}

\keywords{nuclear reactions, nucleosynthesis, abundances -- supernovae: general -- supernovae: individual (SN 2012cg)}


\section{Introduction}
\label{sec:12cg_intro}

The declining light curves of Type Ia supernovae (SNe Ia), as well as some core-collapse SNe, are powered by the radioactive decay of $^{56}{\rm Ni}~\to~^{56}{\rm Co}~\to~^{56}{\rm Fe}$, which releases $\gamma$-rays that thermalize with the expanding ejecta, to be re-emitted in the optical and near-infrared \citep{Truran1967,Colgate1969}. As the ejecta continue to expand, the column density decreases so that the ejecta become progressively more transparent to $\gamma$-rays. However, 19\% of $^{56}{\rm Co}$ decays proceed via positron emission, and after $\sim200$ days these positrons become the dominant energy source and continue to heat the ejecta (e.g., \citealt{1979ApJ...230L..37A,1993ApJ...405..614C,1997A&A...328..203C,1999ApJS..124..503M}). Assuming 100\% trapping of the positrons emitted by these decays, the observed light curves of SNe Ia 300--600 days after explosion (e.g.,  \citealt{2004A&A...428..555S,2006AJ....132.2024L}) are consistent with the decay of $^{56}$Co. However, the apparent slow-down of SN Ia light curves past 800 days, as seen in, e.g., SN 1992A \citep{1997A&A...328..203C}, SN 2003hv \citep{2009A&A...505..265L}, and SN 2011fe \citep{2014ApJ...796L..26K}, indicates that even complete trapping of the $^{56}$Co positrons cannot explain the observed luminosity; additional radioactive heating channels may be required. \citet{2009MNRAS.400..531S} suggested that the slower decays of $^{57}{\rm Co}~\to~^{57}{\rm Fe}$ ($t_{1/2}\approx272$ days) and $^{55}{\rm Fe}~\to~^{55}{\rm Mn}$ ($t_{1/2}\approx1000$ days) will produce internal-conversion electrons, Auger electrons, and X-rays that would largely be trapped, deposit their energy, and slow down the decline of the light curve beginning $\sim800$ days after maximum light.

While there is a wide consensus that SNe Ia are the result of a thermonuclear explosion of a carbon--oxygen white dwarf, it is still unclear how the white dwarf is detonated and consumed (for a recent review, see \citealt{2014ARA&A..52..107M}). \citet{2012ApJ...750L..19R} have used the leptonic energy injection mechanisms due to the decays of $^{57}$Co and $^{55}$Fe described above to provide a testable prediction for two popular SN Ia explosion scenarios: the ``delayed-detonation'' and ``violent merger'' models \citep{1991A&A...245..114K,2011A&A...528A.117P,2012ApJ...747L..10P}. In the first scenario, a white dwarf with a mass close to the Chandrasekhar limit ($M_{\rm Ch}\sim1.4$ M$_{\odot}$) first goes through a subsonic deflagration phase before transitioning to a supersonic detonation. In the second scenario, the white dwarf detonates due to a merger with a second, slightly lower-mass white dwarf, and both stars have sub-Chandra masses (1.1 and 0.9 M$_{\odot}$). Each of these scenarios predicts that different amounts of $^{57}{\rm Co}$ and $^{55}{\rm Fe}$ will be created during the explosion. For example, in the merger scenario, where the densities of the white dwarfs are $<2\times10^8$ g cm$^{-3}$, most of the iron-group elements, including $^{55}{\rm Co}$ (which then decays to the longer-lived $^{55}{\rm Fe}$), are produced in incomplete Si burning and $\alpha$-rich freezeout from nuclear statistical equilibrium. In contrast, the higher density of the delayed-detonation white dwarf leads to ``normal'' freezout (which sets in at $\sim2\times10^8$ g cm$^{-3}$), which produces more $^{55}{\rm Co}$ \citep{1986A&A...158...17T}. The different abundances of $^{55}{\rm Fe}$ and $^{57}{\rm Co}$ then lead to bolometric light curves with appreciably different slopes at $>1200$ days after maximum light.   

\begin{figure*}
 \centering
 \includegraphics[width=\textwidth]{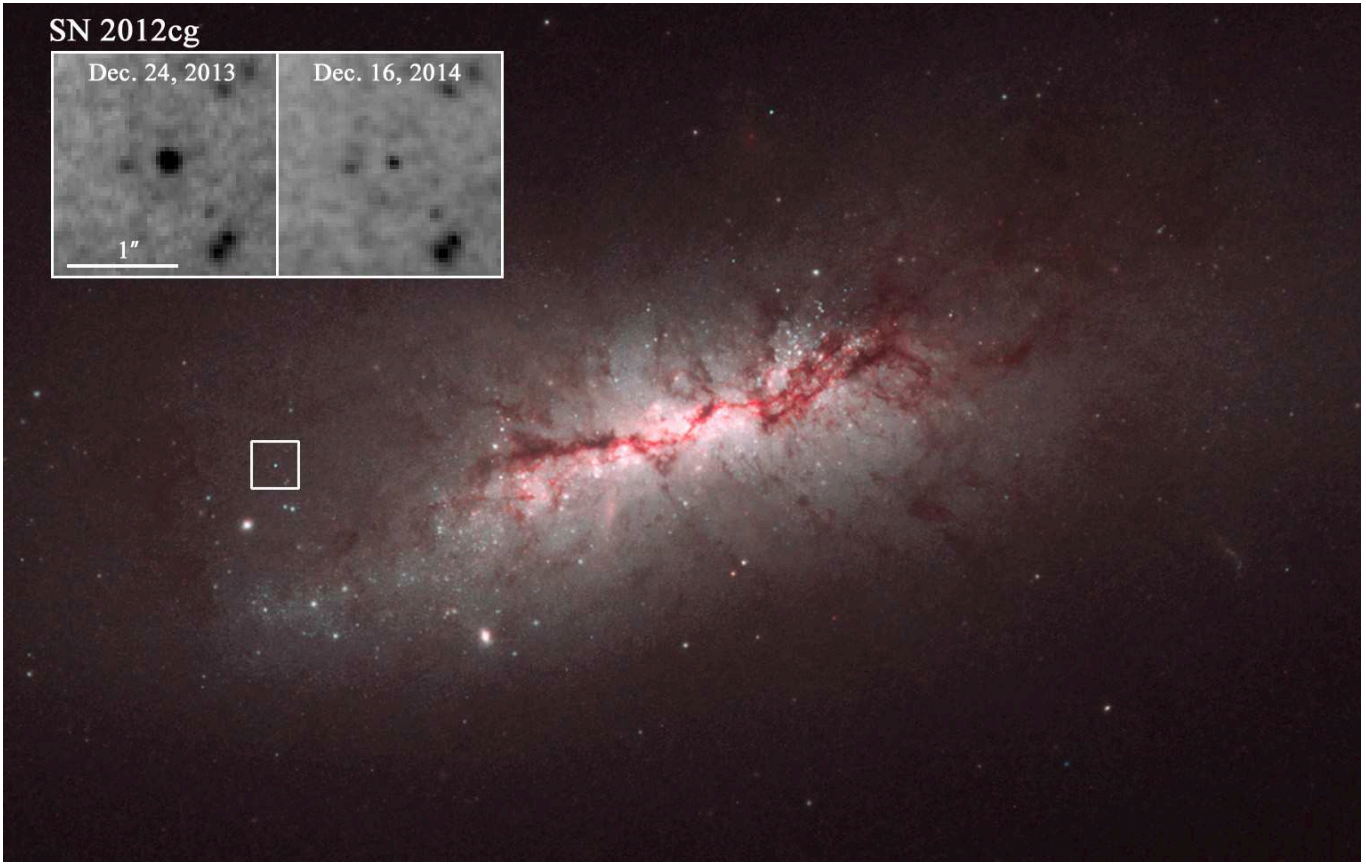}
 \caption{\hst\ color composite of NGC 4424 and SN 2012cg (in the center of the white box). The image is composed of all {\it F160W} (red), {\it F814W} (green), and {\it F555W} (blue) images taken as part of the GO--12880 program. The inset shows two blow-ups of the vicinity of SN 2012cg taken with the {\it F350LP} filter a year apart on 2013 December 24.9 (GO--12880) and 2014 December 16.5 (GO--13799). Each panel is $2^{\prime\prime}$ on a side. The chip gap of the WFC3/UVIS detector has been filled in based on data from the {\it F160W} images. North is up and East is to the left.}
 \label{fig:image}
\end{figure*}

The late-time observations of the core-collapse SNe 1998bw \citep{2009MNRAS.400..531S} and 1987A \citep{2014ApJ...792...10S} were shown to be consistent with the leptonic heating mechanism suggested by \citet{2009MNRAS.400..531S}. When this paper was submitted, this prediction had not been tested for any normal SN Ia (e.g., SN 2001el, observed by \citealt{2007A&A...470L...1S}, as well as the SNe Ia collected by \citealt{2001ApJ...559.1019M} and \citealt{2006AJ....132.2024L}, only go out to $\sim600$ days after explosion). Tantalizingly, the very last data points on the light curves of SN 1992A ($\sim950$ days; \citealt{1997A&A...328..203C}) and SN 2003hv ($\sim700$ days; \citealt{2009A&A...505..265L}) might be showing the beginning of a slow-down of their optical light curves. Very recently, \citet{2015ApJ...814L...2F} published their analysis of the spectrum of SN 2011fe taken 1034 days after explosion by \citet{2015MNRAS.448L..48T}. At this epoch, they find strong evidence of the need for energy injection by $^{57}$Co and, in order to reproduce the observed flux level of the spectrum, require a production ratio of $^{57}{\rm Ni}/^{56}{\rm Ni}$ that is roughly $2.8$ times Solar. Here, we test the predictions made by \citet{2009MNRAS.400..531S} and \citet{2012ApJ...750L..19R} by following the late-time lightcurve of the SN Ia SN 2012cg.

SN 2012cg was discovered on 2012 May 17 (UT) in the nearby spiral galaxy NGC 4424 (at a stellar-kinematics based distance of $15.2\pm1.9$ Mpc; \citealt{2008ApJ...683...78C}) by the Lick Observatory Supernova Search \citep{2001ASPC..246..121F,2012CBET.3111....1K} and spectroscopically classified as a SN Ia by \citet{2012CBET.3111....2C} and \citet{2012CBET.3111....3M}. {\it Hubble Space Telescope} (\hst) pre-explosion images of NGC 4424 revealed no source at the location of the SN down to limits of $M_V\sim-6.0$ and $M_I\sim-5.4$ mags and excluded most supergiants as potential binary companions \citep{2012ATel.4226....1G}. \citet{2012ATel.4453....1C} did not detect the SN in the radio and constrained any possible circumstellar material to either lie at distances $>10^{16}$ cm or be distributed in thin shells $<10^{15}$ cm wide. 


\section{Observations}
\label{sec:12cg_obs}

We imaged SN 2012cg ($\alpha=12^{\rm h}27^{\rm m}12.83^{\rm s}$, $\delta=+09^{\circ}25^{\prime}13.1^{\prime\prime}$) with the \hst\ Wide Field Camera 3 (WFC3) wide-band filters {\it F350LP}, {\it F555W}, {\it F814W}, and {\it F160W} under \hst\ program GO--12880 (PI Riess) on 2013 December 24.9; 2014 January 1.2, 8.8, 17.1, 23.6, and 28.9; February 4.0, 11.8, 20.8, and 27.5; and March 9.6. At these dates the SN was $\approx570$--640 days past maximum light. We subsequently observed SN 2012cg under \hst\ program GO--13799 (PI Graur) with the WFC3 filter {\it F350LP} on 2014 December 16.5, 2015 February 6.9, and 2015 April 26.6, when the SN was $\approx925$--1055 days beyond peak light. Here, we concentrate on observations carried out with the long-pass {\it F350LP} filter. Centered at wavelength $\lambda_0=584.6$~nm and with a width of 475.8~nm, {\it F350LP} is the second-widest WFC3 filter, after {\it F200LP}, covering a wavelength range of $\sim300$--$1000$~nm. 

\begin{table*}
 \centering
 \caption{Late-time photometry of SN 2012cg}
 \begin{tabular}{cccccccccc}
  \hline
  Date & MJD & Phase & Filter & Exposure Time & Magnitude & $\Delta$(Magnitude) & Flux & $\Delta$(Flux) & \hst\ Program \\
   & (days) & (days) &  & (s) &  (AB mag) & (AB mag) & ($\mu$Jy) & ($\mu$Jy) & \\
  \hline
  2013 Dec 24.9 & 56650.9 & 567.9  & {\it F350LP} & 1200 & 23.094 & 0.039 & 2.10 & 0.08 & 12880 \\
  2013 Dec 24.9 & 56650.9 & 567.9  & {\it F555W}  & 936  & 22.750 & 0.019 & 2.88 & 0.05 & 12880 \\
  2014 Jan 01.2 & 56658.2 & 575.2  & {\it F350LP} & 1200 & 23.217 & 0.040 & 1.88 & 0.07 & 12880 \\
  2014 Jan 01.2 & 56658.2 & 575.2  & {\it F814W}  & 900  & 22.867 & 0.034 & 2.59 & 0.08 & 12880 \\
  2014 Jan 08.8 & 56665.8 & 582.8  & {\it F350LP} & 1260 & 23.271 & 0.041 & 1.78 & 0.07 & 12880 \\
  2014 Jan 08.8 & 56665.8 & 582.8  & {\it F160W}  & 906  & 21.35  & 0.14  & 10.5 & 1.3  & 12880 \\
  2014 Jan 17.1 & 56674.1 & 591.1  & {\it F350LP} & 1200 & 23.341 & 0.042 & 1.67 & 0.06 & 12880 \\
  2014 Jan 17.1 & 56674.1 & 591.1  & {\it F160W}  & 906  & 21.51  & 0.17  & 9.0  & 1.4  & 12880 \\
  2014 Jan 23.6 & 56680.6 & 597.6  & {\it F350LP} & 1230 & 23.404 & 0.043 & 1.58 & 0.06 & 12880 \\
  2014 Jan 23.6 & 56680.6 & 597.6  & {\it F814W}  & 900  & 23.173 & 0.046 & 1.95 & 0.08 & 12880 \\
  2014 Jan 28.9 & 56685.9 & 602.9  & {\it F350LP} & 1200 & 23.439 & 0.043 & 1.53 & 0.06 & 12880 \\
  2014 Jan 28.9 & 56685.9 & 602.9  & {\it F814W}  & 938  & 23.291 & 0.050 & 1.75 & 0.08 & 12880 \\
  2014 Feb 04.0 & 56692.0 & 609.0  & {\it F350LP} & 1260 & 23.506 & 0.044 & 1.44 & 0.06 & 12880 \\
  2014 Feb 04.0 & 56692.0 & 609.0  & {\it F160W}  & 906  & 21.80  & 0.21  & 6.9  & 1.3  & 12880 \\
  2014 Feb 11.8 & 56699.8 & 616.8  & {\it F350LP} & 1260 & 23.570 & 0.045 & 1.35 & 0.06 & 12880 \\
  2014 Feb 11.8 & 56699.8 & 616.8  & {\it F160W}  & 906  & 21.79  & 0.21  & 7.0  & 1.4  & 12880 \\
  2014 Feb 20.8 & 56708.8 & 625.8  & {\it F350LP} & 1230 & 23.640 & 0.047 & 1.27 & 0.06 & 12880 \\
  2014 Feb 20.8 & 56708.8 & 625.8  & {\it F555W}  & 900  & 23.132 & 0.026 & 2.03 & 0.05 & 12880 \\
  2014 Feb 27.5 & 56715.5 & 632.5  & {\it F350LP} & 1230 & 23.738 & 0.048 & 1.16 & 0.05 & 12880 \\
  2014 Feb 27.5 & 56715.5 & 632.5  & {\it F814W}  & 900  & 23.566 & 0.063 & 1.36 & 0.08 & 12880 \\
  2014 Mar 09.6 & 56725.6 & 642.6  & {\it F350LP} & 1260 & 23.785 & 0.050 & 1.11 & 0.05 & 12880 \\
  2014 Mar 09.6 & 56725.6 & 642.6  & {\it F160W}  & 906  & 21.96  & 0.24  & 5.9  & 1.3  & 12880 \\
  2014 Dec 16.5 & 57007.5 & 924.5  & {\it F350LP} & 2688 & 25.76  & 0.22  & 0.18 & 0.04 & 13799 \\
  2015 Feb 06.9 & 57059.9 & 976.9  & {\it F350LP} & 2688 & 25.94  & 0.26  & 0.15 & 0.04 & 13799 \\
  2015 Apr 26.6 & 57138.6 & 1055.6 & {\it F350LP} & 2688 & 26.29  & 0.29  & 0.11 & 0.03 & 13799 \\
  \hline
  \multicolumn{10}{l}{\textit{Note.} All photometry is measured in an aperture with a diameter of $0.4^{\prime\prime}$ (except for the final measurement on 2015 26.6, for which we} \\
  \multicolumn{10}{l}{used an aperture with a diameter of $0.24^{\prime\prime}$). Phases are relative to $B$-band maximum light on 2012 June 4.5. The measurements here have} \\
  \multicolumn{10}{l}{not been corrected for reddening by either Galactic or host-galaxy dust.}
 \end{tabular}
 \label{table:mags}
\end{table*}

In Figure~\ref{fig:image}, we show a color composite of NGC 4424 and SN 2012cg along with an inset that shows {\it F350LP} observations of the SN taken a year apart. While the SN has faded appreciably between the two visits, it is still visible $>900$ days after explosion. We present our measurements of the light curve of SN 2012cg in Figure~\ref{fig:lightcurve}. 


\section{Analysis}
\label{sec:12cg_analysis}

We performed aperture photometry of SN 2012cg in a $0.4^{\prime\prime}$-diameter aperture using the hstsnphot\footnote{\href{https://github.com/srodney/hstsntools}{https://github.com/srodney/hstsntools}} program used by the CLASH and CANDELS SN surveys \citep{Graur2014,2014AJ....148...13R}, which is based on the IRAF\footnote{\href{http://iraf.noao.edu/}{http://iraf.noao.edu/}} {\it apphot} routine \citep{1986SPIE..627..733T}. Due to the faintness of the SN on the last observation epoch, 2015 April 26.6, we used a smaller aperture with a diameter of $0.24^{\prime\prime}$, along with a larger aperture correction, to capture the low flux counts. Zero points and aperture corrections were measured by the Space Telescope Science Institute.\footnote{\href{http://www.stsci.edu/hst/wfc3/phot\_zp\_lbn}{http://www.stsci.edu/hst/wfc3/phot\_zp\_lbn}} Our photometry of SN 2012cg, uncorrected for Galactic or host-galaxy reddening, is reported in Table~\ref{table:mags}.

We correct our measurements for both Galactic extinction along the line of sight to SN 2012cg and the extinction caused by dust in the SN host galaxy. The Galactic line-of-sight extinction toward SN 2012cg in {\it F350LP}, {\it F555W}, {\it F814W}, and {\it F160W} is 0.054, 0.059, 0.032, and 0.011 mag, respectively \citep{2011ApJ...737..103S}. By using the \citet{1989ApJ...345..245C} reddening law, an $E(B-V)=0.18$ mag reddening \citep{2012ApJ...756L...7S,2015arXiv150707261M}, and assuming $R_V=3.1$, we calculate host-galaxy extinctions of 0.515, 0.582, 0.333, and 0.114 mag in the same filters.

SN 2012cg peaked in the $B$ band sometime during 2012 June 2--5, at which point it reached an absolute magnitude similar to that of SN 2011fe (as measured by \citealt{2013NewA...20...30M}). While it is similar to SN 2011fe in many respects (e.g., in its color evolution around peak, as well as the apperance of its nebular spectra; see \citealt{2015MNRAS.453.3300A}, but cf. \citealt{2015arXiv150707261M} for a possible early excess of blue light. For more information on SN 2011fe, see, e.g., \citealt{Li2011fe,2013PASA...30...46C,KasenNugentSN2011fe,2014MNRAS.442L..28G}), it has a slightly broader light curve. Specifically, by fitting early-time data, \citet{2012ApJ...756L...7S} measured $t_B^{max} = 2012~{\rm June}~2.0\pm0.75$, $M_B = -19.73\pm0.30$ mag, and a $B$-band light curve that was slightly \emph{narrower} than SN 2011fe (the only work to reach such a conclusion). \citet{2013NewA...20...30M} measured $t_B^{max} = 2012~{\rm June}~ 4.5$, $M_B=-19.55$ mag, $\Delta m_{15,B} = 1.039$ (compared to 1.108 for SN 2011fe), and a reconstructed bolometric light curve that reaches a similar (though slightly higher) peak luminosity as SN 2011fe, with a slightly broader shape overall (see their figure 4). \citet{2015arXiv150707261M} measured an earlier $t_B^{max} = 2012~{\rm June}~3.3\pm0.5$, $M_B=-19.62\pm0.08$ mag, and $\Delta m_{15,B} = 0.86\pm0.02$. Finally, \citet{2015MNRAS.453.3300A} found that $t_B^{max} = 2012~{\rm June}~2.6\pm0.3$ and $s_B = 1.12\pm0.02$, where the latter is the \citet{1997ApJ...483..565P} stretch parameter. Here, we adopt the value of peak $B$-band light derived by \citet{2013NewA...20...30M}, so that all phases are reported relative to 2012 June 4.5.

The decay of $^{56}$Co to $^{56}$Fe has a half-life time of $t_{1/2}=77.2$ days \citep{Junde20111513}. Assuming that the shape of the light curve is set by $^{56}$Co decay, we fit an exponential function to the {\it F350LP} measurements at $570<t<640$ days and find a $^{56}$Co half-life of $82.7\pm5.3$ days with a $\chi^2$ value of $1.7/9$ for nine degrees of freedom. A similar fit to the {\it F160W} data results in a half-life of $69\pm26$ days with $\chi^2=0.6/3$. These values are consistent with the half-life time obtained in the laboratory, and with each other. However, they also hint that at $>500$ days, we may be seeing a redistribution of light from the infrared to the optical, as previously seen in observations of SN 1992A and SN 2003hv (though based on only one and two measurements, respectively; \citealt{1997A&A...328..203C,2009A&A...505..265L}). 

\begin{figure}
 \centering
 \includegraphics[width=0.475\textwidth]{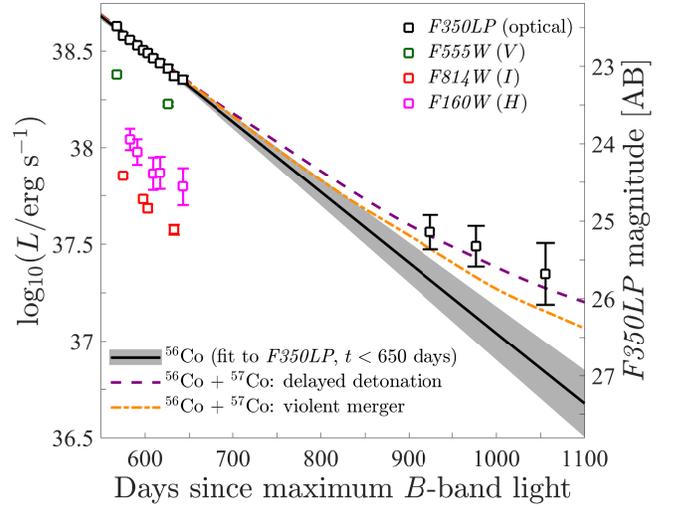}
 \caption{Light curve of SN 2012cg at $570<t<1055$ days after $B$-band maximum light. {\it F350LP, F555W, F814W}, and {\it F160W} luminosities appear as white, green, red, and magenta squares, as noted. The solid black curve is the best-fitting exponential function to the {\it F350LP} measurements in the range $570<t<640$ days. The shaded region around this curve represents the confidence region of the fit given the 68\% statistical uncertainties of the fitted parameters. The dashed purple and dotted-dashed orange curves are the predicted bolometric luminosities of SN 2011fe for the delayed-detonation and violent merger progenitor scenarios, respectively, scaled down so that the predicted $^{56}$Co-only decay matches the fit to the $570<t<640$ days {\it F350LP} data. The {\it F350LP} measurements at $t>900$ days indicate that the light curve is no longer dominated solely by the decay of $^{56}$Co; instead, its decline is being slowed down by light emitted from the decay of $^{57}$Co to $^{57}$Fe.}
 \label{fig:lightcurve}
\end{figure}

In Figure~\ref{fig:lightcurve}, we extrapolate the {\it F350LP} fit out to 1100 days to represent the expected light curve due solely to the decay of $^{56}$Co to $^{56}$Fe. The three {\it F350LP} measurements at $t>900$ days are brighter than the extrapolated fit by factors of $1.8$--$3.2$. To ascertain at what significance these measurements are inconsistent with the null hypothesis that the light curve is due solely to $^{56}$Co decay, we compare between the three $t>900$ days measurements and the model, while taking into account the uncertainties of both. When fitting these measurements, the values for the index and scaling of the decaying exponential function that resulted from fitting the $570<t<640$ days measurements receive a $\chi^2$ value of $9.9/1$, which calls for a rejection of the null hypothesis at a significance of $>99.8$\% (i.e., a $p$-value of $\sim0.002$, comparable to a $3\sigma$ detection).

Next, we test whether the $t>900$ days {\it F350LP} measurements are consistent with the predictions for the delayed-detonation and violent merger progenitor scenarios. The original predictions are bolometric light curves of SN 2011fe, while we measure the optical component of the light curve of SN 2012cg captured by the {\it F350LP} filter. Although the predictions were made for a different SN Ia, SN 2012cg was observed to have a similar intrinsic maximal brightness as SN 2011fe but a slightly broader light curve. We solve both of these potential difficulties by comparing between the predicted bolometric light curve due to $^{56}$Co decay for SN 2011fe and our fit to the $570<t<640$ days {\it F350LP} observations of SN 2012cg and scaling down the predicted bolometric light curves accordingly. The {\it F350LP} measurements are consistent with the resultant scaled-down predictions but are not precise enough to discriminate with certainty between the two progenitor scenarios.

Here, we have assumed that the emission in {\it F350LP} scales with the bolometric luminosity. \citet{2001ApJ...559.1019M} have shown that at $50<t<600$ days, a constant fraction of the luminosity is emitted in the $V$ band, which dominates the 3500--9700 \AA{} wavelength range (the same range spanned by {\it F350LP}). Past $\sim500$ days, the $V$ band becomes dominated by Fe lines (see \citealt{2015MNRAS.448L..48T}), which, in the absence of the infrared catastrophe (cf. \citealt{2015ApJ...814L...2F}), are the primary coolant of the SN ejecta and hence also proportional to the total luminosity of the SN. 

As noted above, the two progenitor scenarios tested here differ, among other things, in the amounts of radioactive iron-group elements produced during the explosion. Thus, a measurement of the ratio between the amount of $^{56}$Ni and $^{57}$Ni produced during the explosion would provide a strong constraint for any progenitor and explosion model. Here, we compute the luminosity contributed by the decays of $^{56}$Co, $^{57}$Co, and $^{55}$Fe by using the solution to the Bateman equation in the following form (following \citealt{2014ApJ...792...10S}):
\begin{equation}\label{eq:lum}
 L_A(t) = 2.221 \frac{B}{A} \frac{\lambda_A}{\rm days^{-1}} \frac{M(A)}{M_\odot} \frac{q_A}{{\rm keV}} {\rm exp}(-\lambda_A t) \times10^{43}~{\rm erg~s^{-1}},
\end{equation}
where $B=0.235$ is the scaling factor described above; $A$ is the atomic number of the decaying nucleus; $\lambda_A$ is the inverse of the half-life time of the decay chain; $q_A$ is the average energy per decay carried by charged leptons and X-rays; $t$ is the time since explosion; and we fit for the masses, $M(A)$, of $^{56}$Co and $^{57}$Co. Our late-time measurements are not precise enough to also fit for the mass of $^{55}$Fe, so we set its mass by using a ratio of $M(^{57}{\rm Co})/M(^{55}{\rm Fe})\approx0.8$ \citep[model rpc32;][]{2014A&A...572A..57O}. The values of $\lambda_A$ and $q_A$ used here are enumerated in table 1 of \citet{2009MNRAS.400..531S} and table 2 of \citet{2014ApJ...792...10S}. For $^{57}$Co, we also take into account the energy emitted by the decay through the 14.4 keV line (1.32 keV per decay).

Several assumptions go into Equation~\ref{eq:lum}:
\begin{enumerate}
 \item $\gamma$-rays are essentially free-streaming and their contribution to the heating is negligible.
 \item Positrons, electrons, and X-rays are fully trapped and instantaneously deposit all their energy.
 \item The luminosity is equal to the rate of energy deposition at every time, i.e., we do not take ``freeze-out'' into account (see, e.g., \citealt{1998ApJ...496..946K}).    
\end{enumerate}
While these assumptions clearly constitute approximations, we presume they are applicable and justified since:
\begin{enumerate}
 \item The dominant opacity to $\gamma$-rays is Compton scattering. For homologous expansion, the column density, and therefore the opacity to $\gamma$-rays, decreases as $t^{-2}$. Since, e.g., \citet{1997A&A...328..203C} find that for a generic SN Ia model the trapping fraction of $\gamma$-rays at $t=200$ days is already down to 0.5\%, we conclude that the contribution of $\gamma$-rays is probably safely negligible past $\sim600$ days (with the exception of the very low-energy 14.4 keV $\gamma$-ray produced by $^{57}$Co decay, which we include in our analysis). 
 \item The calculations of \citet{2015MNRAS.447.1484S} demonstrate that at $\sim1000$ days full trapping of X-rays produced in radioactive decay of Fe-group nuclei should be a good approximation. The possible escape of positrons is more uncertain. However, at present there is no real indication for positron escape in normal SNe Ia. On the contrary, recent analyses of very late SN Ia observations support full trapping of positrons even past 800 days (see, e.g., \citealt{2009A&A...505..265L,2014ApJ...796L..26K,2015ApJ...814L...2F}). If positrons are fully trapped, the much lower-energy electrons would be trapped as well. 
 \item \citet{2015ApJ...814L...2F} demonstrated that, for the case of the similar SN 2011fe, freeze-out effects are essentially negligible out to $\sim900$ days in the $V$ band, which dominates the emission in {\it F350LP}, taking the \citet{2015MNRAS.448L..48T} SN 2011fe spectrum at 1034 days past explosion as our best guess for the spectral-energy distribution. Furthermore, their freeze-out corrections remain smaller than our measurement uncertainties even at $\sim1050$ days. Freeze-out effects, therefore, constitute a small correction that would not change our conclusions qualitatively.
\end{enumerate}

\begin{figure}
 \centering
 \includegraphics[width=0.475\textwidth]{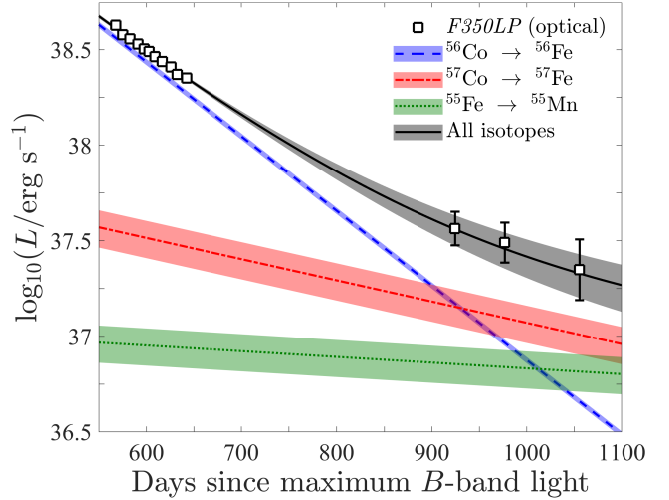}
 \caption{Luminosity contributions from the decays of $^{56}$Co (blue dashed), $^{57}$Co (red dotted-dashed), and $^{55}$Fe (green dotted). The total luminosity produced by these decay chains (black solid) fits the {\it F350LP} measurements with a $\chi^2$ value of 2.1 for 12 degrees of freedom.}
 \label{fig:bateman}
\end{figure}

With a $\chi^2$ value of 2.1 for 12 degrees of freedom, we find a best-fitting $^{56}$Co mass of $M(^{56}{\rm Co})\approx0.7~M_\odot$ and a mass ratio of $M(^{57}{\rm Co})/M(^{56}{\rm Co}) = 0.043^{+0.012}_{-0.011}$. We show the luminosity contributions from each decay chain in Figure~\ref{fig:bateman}. Restricting the fit to account only for $^{56}$Co and $^{57}$Co results in the same $^{56}$Co mass but a mass ratio of $M(^{57}{\rm Co})/M(^{56}{\rm Co}) = 0.068^{+0.019}_{-0.018}$, larger by a factor of $\sim 1.6$ (with a $\chi^2$ value of $2.5/12$). We do not quote an uncertainty for the $^{56}$Co mass as its value is degenerate with the scaling factor $B$. Without a full bolometric light curve, we cannot measure this quantity precisely, but note that it is consistent with $^{56}$Ni masses measured in other SNe Ia \citep{2014MNRAS.445.2535S}. The mass ratio, however, is independent of the scaling factor. In Section~\ref{subsec:12cg_yields}, below, we examine the value of the mass ratio in the context of expected yields from various SN Ia explosion models.

Here, we have used $R_V=3.1$ to correct for the host-galaxy reddening suffered by SN 2012cg. However, based on the observed colors of SN 2012cg around maximum light, \citet{2015MNRAS.453.3300A} find a lower value of $R_V=2.7^{+0.9}_{-0.7}$. While this value is consistent with the one we use here, we note that the value of $R_V$ has no effect on the value of $M(^{57}{\rm Co})/M(^{56}{\rm Co})$, as it depends only on the slope of the {\it F350LP} photometry, not their absolute values.

\section{Discussion}
\label{sec:12cg_discuss}

In this section, we first place the $M(^{57}{\rm Co})/M(^{56}{\rm Co})$ mass ratio found above in the context of current SN Ia explosion models. Then, we address two other possible explanations for the slow-down of the decline of the light curve at $>900$ days: contamination by light echoes or the light from a surviving binary companion. We show the effects of these possible contaminants in Figure~\ref{fig:le}.

\subsection{Expected $^{57}$Co yields from SN Ia explosion models}
\label{subsec:12cg_yields}

Here, we assume that nearly all of the mass of the iron-group elements produced by the explosion originates in the decay of various Ni isotopes, so we can approximate $M(^{57}{\rm Ni})/M(^{56}{\rm Ni}) = M(^{57}{\rm Co})/M(^{56}{\rm Co}) = M(^{57}{\rm Fe})/M(^{56}{\rm Fe})$ \citep{Truran1967,1973ApJS...26..231W}. Under this approximation, the mass ratio we measure here is roughly twice the corresponding $^{57}$Fe/$^{56}$Fe ratio measured for the Sun \citep{2009ARA&A..47..481A}. This value is slightly higher than the predictions of near-$M_{\rm Ch}$ explosion models: W7 of \citet{1999ApJS..125..439I} predicts 1.7 times Solar, the pure turbulent deflagration models of \citet{2014MNRAS.438.1762F} predict $\sim1.5$ times Solar, the delayed-detonation models of CO WDs of \citet{2013MNRAS.429.1156S} predict $\sim1.3$ times Solar, the delayed-detonation models of WDs with carbon-depleted cores of \citet{2014A&A...572A..57O} predict $\sim1.1$ times Solar, and the gravitationally confined detonation models of \citet{2009ApJ...693.1188M} predict 0.8--1.2 times Solar. 

There is a trend that models with larger progenitor metallicity (see \citealt{2013MNRAS.429.1156S}) and models that experience enhanced in-situ neutronization (e.g., W7) yield higher $^{57}$Ni/$^{56}$Ni ratios. This is naturally explained by the fact that $^{57}$Ni is a neutron-rich isotope (28 protons, 29 neutrons) and its production relative to the self-conjugate $^{56}$Ni (28 protons, 28 neutrons) therefore tends to be enhanced in neutron-rich environments. Model yields where the primary WD is significantly less massive than $M_{\rm Ch}$ predict even lower $^{57}$Ni/$^{56}$Ni ratios: the merger model of a $1.1~M_\odot$ CO WD with a $0.9~M_\odot$ CO WD of \citet{2012ApJ...747L..10P} predicts $\sim1.1$ times Solar and the pure detonation models of CO and ONe WDs of \citet{2015A&A...580A.118M} predict around 0.3--0.5 times Solar. 

\citet{2013A&A...557A...3P} have found that the $^{57}$Ni/$^{56}$Ni production ratio is very robust (at the 10\% level) to individual changes of nuclear reaction rates by a factor of 10; rate uncertainties are therefore unlikely to solely account for the difference. Therefore, the high $^{57}$Co/$^{56}$Co ratio we find appears to point toward a near-$M_{\rm Ch}$ progenitor. This is in agreement with \citet{2015arXiv150707261M}, who interpret the observed early UV excess of SN 2012cg as evidence that the SN occurred as an explosion of a near-$M_{\rm Ch}$ WD in the single-degenerate channel.

\subsection{Contamination by a light echo?}
\label{subsec:12cg_le}

Light echoes have been found for other SNe Ia at similar phases hundreds of days after their explosion, e.g., SN1991T \citep{1994ApJ...434L..19S,1999ApJ...523..585S}, SN1995E \citep{2006ApJ...652..512Q}, SN1998bu \citep{2001ApJ...549L.215C}, SN2006X \citep{2008ApJ...677.1060W}, SN2007af \citep{2015ApJ...805...71D}, and SN2014J \citep{2015ApJ...804L..37C}. Typically, these light echoes were between 10--12 mag fainter than the SNe at peak. Due to observational bias, these observed light echoes are at the bright tail of the distribution of expected light-echo brightnesses \citep{2005MNRAS.357.1161P}. Therefore, the excess flux of SN 2012cg at $>900$ days, which is $\sim14$ mag fainter than the SN at peak, is consistent with a light echo. In the following, we investigate this possibility in more detail.

There is no evidence of a resolved light echo at the location of SN 2012cg in the three $>900$-day epochs. Using Dolphot,\footnote{\href{http://americano.dolphinsim.com/dolphot/}{http://americano.dolphinsim.com/dolphot/}} a modified version of HSTphot \citep{2000PASP..112.1383D}, we subtracted a model of the {\it F350LP} point-spread function (PSF) from the location of the SN. In each epoch, a visual inspection of the residual map produced by Dolphot revealed no resolved features. To quantify this statement, we compared the root-mean-square (RMS) of the pixel values at the location of the PSF-subtracted SN, measured in rectangular apertures of $7\times7$, $5\times5$, and $3\times3$ pixels$^2$, to the distribution of RMS values measured in 100 apertures at random locations up to 15 pixels away in any direction from the SN. In each of the three epochs, the RMS value was within the 16th and 84th percentiles of the distribution, i.e., consistent with the background noise. 

\begin{figure}
 \centering
 \includegraphics[width=0.475\textwidth]{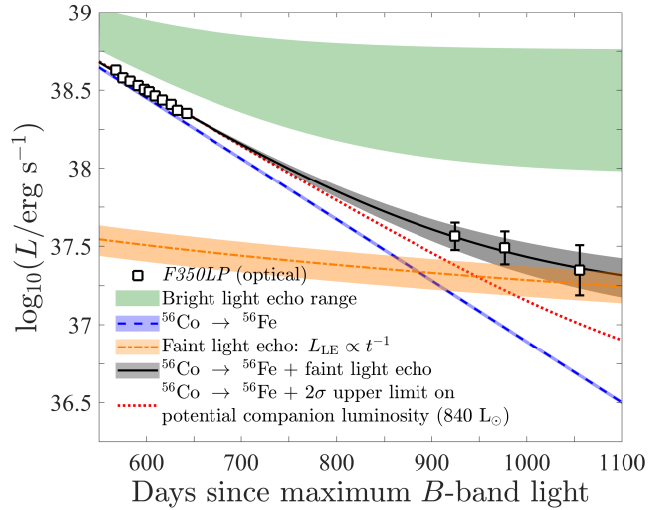}
 \caption{Possible contaminants of the light curve of SN 2012cg. The expected luminosity range of a bright light echo is shown as a green patch, while $^{56}$Co decay (blue dashed) and a faint light echo that decays as $t^{-1}$ (orange dotted-dashed) can combine (black solid) to fit the {\it F350LP} measurements (white squares). The dotted red curve shows the maximal contribution to the SN light from a red-giant or $\sim7~M_\odot$ main-sequence binary companion, according to $2\sigma$ upper limits on the luminosity of such a companion measured from \hst\ pre-explosion images.}
 \label{fig:le}
\end{figure}

We now address the possibility of a non-resolved light echo within the PSF. Let us first consider the shape of the light curve. As we have argued above, the observed excess flux surface brightness is well within the range expected for light echoes. With the reasonable assumption that the scattering dust structures do not vary significantly on the length scales associated with the light echo, the surface brightness of light echoes decays as $t^{-2}$ (e.g., \citealt{1986ApJ...308..225C,2003AJ....126.1939S}). As the light echo is not resolved, we must also take into account that its angular size, which is proportional to the circumference, increases with time, so that the light echo should fade as $t^{-1}$. We test this hypothesis by fitting a combination of $^{56}$Co decay and a light echo decaying as $t^{-1}$ to the data, with the light echo normalization set as a free parameter. Figure~\ref{fig:le} shows that this model, with $\chi^2 = 1.8/12$, fits the data well.

Next, we consider whether the observed colors are consistent with the light echo above. Light-echo flux is dominated by the light of the SN around peak, so to first order the color of the light echo is similar to the color of the SN at peak. In addition, the scattering by the dust favors blue light, which can shift the color of the light curve bluewards by a few tenths of a magnitude, depending on the scattering angle (e.g., \citealt{2012PASA...29..466R}). 

At $0.3$ days, \citet{2015MNRAS.453.3300A} measure $V-i^\prime=-0.220\pm0.045$ mag (Vega, corrected for Galactic extinction). Our photometry allows us to measure the {\it F555W}$-${\it F814W} color (similar to $V-I$) of SN 2012cg at 570--640 days. As our {\it F555W} and {\it F814W} measurements were not taken on the same days, we use the python AstroML package \citep{astroMLText}\footnote{\href{http://www.astroml.org/}{http://www.astroml.org/}} to perform Gaussian process regression and measure the color at any phase within this range. At 600 days, we measure ${\it F555W-F814W}=0.140\pm0.055$ mag (Vega, corrected for Galactic extinction). Hence, at 600 days, SN 2012cg was redder than it was at peak. However, at 600 days there is no excess flux, assuming the light curve is powered solely by $^{56}$Co decay (see Figure~\ref{fig:lightcurve}). The combination of $^{56}$Co decay and a faint light echo shown in Figure~\ref{fig:le} would mean a flux excess of 11\%, relative to the $^{56}$Co component. Therefore, any contamination by a faint light echo would only minimally change the observed combined color. Thus, the observed color of SN 2012cg at 600 days is consistent with either no light echo or a light echo faint enough that it leaves no trace at this phase. At $>900$ days, where the excess flux dominates the observed flux and the light-echo hypothesis could be tested best with the color, we only have observations in one filter. 

Without much fine-tuning, a light echo can explain the observed colors and light-curve shape. Thus, with no spectrum or color information at late times, we cannot conclusively rule out contamination of our measurements by a light echo. It is a coincidence of nature that the functional forms of the light-echo and $^{57}$Co-decay luminosities exhibit the same behavior in the time range probed here, as can be seen by comparing Figures~\ref{fig:bateman} and \ref{fig:le}. Unfortunately, further observations of SN 2012cg cannot inform this issue. Our final {\it F350LP} measurement already had a signal-to-noise ratio of 3.3; by now, the SN should be as faint or fainter than the background light of NGC 4424.

\subsection{Contamination by a surviving binary companion?}
\label{subsec:12cg_SD}

To test whether our measurements at $t>900$ days indicate that the SN had faded to the point where a surviving binary companion could shine through, we first fit all {\it F350LP} measurements with a combination of an exponential function, for the SN light curve, and a constant luminosity from the presumptive companion. The best-fitting companion has a luminosity of $(4.4^{+0.8}_{-1.4})\times10^3~L_\odot$ ($\chi^2=2.4/12)$, consistent with a red giant. However, pre-explosion \hst\ Wide-Field Planetary Camera 2 images of the location of the SN have produced $2\sigma$ upper limits of ${\it F606W}=25.5$ and ${\it F814W}=25.8$ mag \citep{2012ATel.4226....1G}. Taking into account Galactic and host-galaxy extinctions for these filters results in a $2\sigma$ upper limit on the {\it F350LP} luminosity of a companion star of $\ltsim840~L_\odot$, as shown in Figure~\ref{fig:le}. Such a luminosity is consistent with either a red giant or a $\sim7~M_\odot$ main-sequence star, but constitutes only $\sim10$\% of our measured {\it F350LP} luminosities at $t>900$ days. Our upper limit is consistent with the conclusion of \citet{2015arXiv150707261M} that the pre-maximum UV excess observed in SN 2012cg is the possible signature of interaction between the SN ejecta and a $\sim6~M_\odot$ main-sequence companion. We conclude that, while we cannot rule out the presence of a surviving companion star, even if it is present, its contribution to the light curve is still negligible.

\section{Conclusions}
\label{sec:12cg_conclude}

In conclusion, observations of the SN Ia SN 2012cg more than 900 days after its explosion allow us to reject, at a significance of $>99.8$\% (i.e., $3\sigma$), the null hypothesis that the light curve of this SN Ia is powered solely by the radioactive decay chain $^{56}{\rm Ni}~\to~^{56}{\rm Co}~\to~^{56}{\rm Fe}$. Instead, we have shown that our measurements are consistent with predictions that at these late times, the light curve becomes dominated by reprocessing of internal-conversion and Auger electrons, as well as X-rays, emitted by the decay of $^{57}$Co into $^{57}$Fe, as predicted by \citet{2009MNRAS.400..531S}. This constitutes the first evidence for the synthesis of radioactive $^{57}$Co during a SN Ia explosion. 

We have tested whether the late-time behavior of the light curve of SN 2012cg could be due to contamination by either a light echo or the luminosity from a surviving binary companion. Pre-explosion imaging sets an upper limit on the latter, which makes it a negligible source of contamination. A combination of $^{56}$Co decay and a faint light echo, $\sim14$ mag fainter than SN 2012cg at peak, fits the data well. Unfortunately, without spectra or colors at late times, we cannot rule out the possibility of light-echo contamination. 

Assuming no contamination by light echoes, we have measured a ratio of $0.043^{+0.012}_{-0.011}$ between the masses of $^{57}$Co and $^{56}$Co produced by the decay of Ni isotopes synthesized in the explosion. This ratio, which is roughly twice the corresponding Solar $^{57}$Fe/$^{56}$Fe ratio, implies that the progenitor of this SN was probably a near-$M_{\rm Ch}$ WD. After this paper was submitted, \citet{2015ApJ...814L...2F} found that a ratio $\sim2.8$ times the Solar value was required to fit a spectrum of SN 2011fe $\sim1000$ days after explosion. It remains to be seen whether observations of other SNe Ia at such late times provide similar ratios and whether SN Ia light curves $>2000$ days after explosion will reveal conclusive evidence of similar leptonic energy-injection channels from the decay of the longer-lived $^{55}$Fe.


\begin{acknowledgements}
We thank Federica Bianco, Saurabh Jha, Wolfgang Kerzendorf, Robert Kirshner, Curtis McCully, Xiangcun Meng, Maryam Modjaz, Dovi Poznanski, and the anonymous referee for helpful discussions and comments. We also thank Rahman Amanullah, Melissa Graham, Kate Maguire, and Jeffrey Silverman for sharing with us their spectra of SNe 2011fe and 2012cg; and Lisa Frattare (STScI) for creating the color composite image of NGC 4424. IRS was supported by the ARC Laureate Grant FL0992131. This work is based on data obtained with the NASA/ESA {\it Hubble Space Telescope}. Support for Programs GO--12880 and GO--13799 was provided by NASA through grants from the Space Telescope Science Institute, which is operated by the Association of Universities for Research in Astronomy, Incorporated, under NASA contract NAS5-26555. This research has made use of NASA's Astrophysics Data System (ADS) Bibliographic Services and of the NASA/IPAC Extragalactic Database (NED), which is operated by the Jet Propulsion Laboratory, California Institute of Technology, under contract with NASA. The {\it HST} imaging data used in this paper can be obtained from the Barbara A. Mikulski Archive for Space Telescopes (MAST) at https://archive.stsci.edu. The ESA/ESO/NASA FITS Liberator was used to create Figure~\ref{fig:image}.

\end{acknowledgements}

{\it Facility:} \facility{HST (WFC3)}
\smallskip



\end{document}